\documentstyle[prl,aps,twocolumn,epsf]{revtex}
\begin{document}
\title{Scattering matrices with block symmetries}
\author{Karol \.Zyczkowski}
\address{Instytut Fizyki M.~Smoluchowskiego, Uniwersytet Jagiello\'nski,
ul. Reymonta 4, 30-059 Krak\'ow, Poland}
\maketitle
\begin{abstract}
Scattering matrices with block symmetry, which corresponds to
scattering process on cavities with geometrical symmetry,
are analyzed. The distribution of transmission coefficient
is computed for different number of channels in the case of a system
with or without the time reversal invariance.
An interpolating formula for the case of
gradual time reversal symmetry breaking is proposed.
\end{abstract}

\section{Introduction}

Ensembles of random matrices introduced
in context of the theory of nuclear spectra
by Dyson \cite{dys63}  long time ago
found a novel application in problems of
chaotic scattering \cite{mps85,bs88,lewd91,bm94,jpb94,bro95,szz96}.
The $S$--matrix corresponding to time reversal invariant problems
can be described by matrices of
circular orthogonal ensemble (COE), while the
circular unitary ensemble (CUE) is applicable
if no antiunitary symmetry exists.

The process of chaotic scattering in cavities with a geometrical
 symmetry
can be represented by an $S$-matrix with block symmetry. Such
matrices where recently introduced by
Gopar et al. \cite{GMMB96},
who discussed both cases: with and without
time--reversal symmetry. They found a link between
the invariant measure of such ensembles
and the measures of canonical circular ensembles and computed the
distribution of transmission coefficient for one and two incoming
channels.

In this Brief Report we generalize the results of Gopar et al.
\cite{GMMB96}. Using an idea of composed ensembles of random matrices
we simplify analytical
calculation of the transmission coefficient $T$. This method allows us
to obtain the distributions $P(T)$ for interpolating ensembles
corresponding to breaking of the time reversal symmetry
and to treat the case of large number of channels.
Generating random unitary matrices of interpolating ensembles
according to the method described in \cite{ZK96}
we verify numerically proposed distributions.

\section{Block symmetric scattering matrices}
We introduce the random S--matrices with block symmetry following the
method of Gopar et al. \cite{GMMB96}.
The $2M \times 2M$ scattering matrix possess the structure
\begin{equation}
S = \left(\matrix{r  & t  \cr t  & r\cr}\right),
\label{S1}
\end{equation}
where $M\times M$ matrices $r$ and $t$ describe reflection and
transmission processes, respectively. S--matrix
can be brought into a block diagonal form
\begin{equation}
S' = R_0S R_0^T \left(\matrix{s_1  & 0  \cr 0  & s_2\cr}\right),
\label{S2}
\end{equation}
  where $s_1=r+t$ and $s_2=r-t$ are unitary matrices and
 $R_0$ stands for rotation matrix consisting of
M-dimensional unit matrices $1_M$
\begin{equation}
 R_0 = {1 \over \sqrt{2}} \left(\matrix{1_M  & 1_M  \cr -1_M  &
1_M\cr}\right).
\label{S3}
\end{equation}
Of great physical importance is the total transmission coefficient
\begin{equation}
 T={\rm tr} (tt^{\dagger}),
\label{T1}
\end{equation}
which is proportional to the conductance $G$ of the cavity,
$G=(2e^2/h)T$. Since $t=(s_1-s_2)/2$, the transmission coefficient reads
\begin{equation}
 T={M \over 2} -{1\over 2} {\rm Re[tr} (s_1s_2^{\dagger})].
\label{T2}
\end{equation}
Following Gopar et al. \cite{GMMB96} we assume that unitary matrices
$s_1$ and
$s_2$ are statistically independent and pertain to the same universality
class. For canonical circular ensembles of random matrices the
joint probability distribution (JPD) of eigenphases is
given by a single formula \cite{mehta}
\begin{equation}
P_{U_{\beta}}[\varphi_1,\dots,\varphi_M ]=C_{\beta,M} \prod_{i>j}
|e^{i\varphi_i}-e^{i\varphi_j}|^{\beta},
\label{P1}
\end{equation}
where $C_{\beta,M}$ stands for normalization constant, while
$U_{\beta}$ represents
Poissonian, orthogonal and unitary circular ensemble for
$\beta$ equal to $0,1$ and $2$, respectively.

Formula (\ref{T2}) contains a product of two unitary matrices
$s_1s_2$ (one can redefine the second matrix writing
$s_2'=s_2^{\dagger}$). Further calculation base on the relation linking
the JPD of eigenvalues of the product of two unitary
matrices drown independently from any canonical ensemble
\begin{equation}
  P_{U_{\beta}\times U_{\beta}} [\varphi_1,\dots,\varphi_M ] =
P_{U_{\beta}}[\varphi_1,\dots,\varphi_M ]
\label{P2}
\end{equation}
The left hand side of the above equation formally represents
the JPD characteristic to spectra of a composed ensemble defined
via product of two random matrices, each
specified by a certain probability distribution.
 For unitary ensemble ($\beta=2$) this
equality follows from the invariance properties of CUE, which
corresponds to the Haar measure on the unitary
group. The same concerns the case $\beta=0$, since the circular
Poissonian ensemble (CPE)
can be defined by the Haar measure in the space of diagonal unitary
matrices.
In the case of symmetric unitary matrices
($\beta=1$) this property follows from the definition of COE,
which is invariant with respect to transformations
$s\to s'=XsX^T$ \cite{dys63}, where $X$ denotes any unitary matrix.
Analyzed matrix $s_1s_2$ is similar to
$s_1^{1/2}s_2s_1^{1/2}$. Since $s_1$ is symmetric,
so is $s_1^{1/2}$, which may play the role of $X$ in the invariance
condition.  The JPD of eigenvalues
of $s_1s_2$ (averaged over the composed ensemble)
is thus the same as this of of $s_2$ and is
given by  (\ref{P1}) with $\beta=1$.
In spite of this result
the measure of the composed ensemble containing products of
two symmetric unitary matrices differs form this characteristic of COE
 \cite{KPZ96}.

Since the JPD of eigenvalues does not define the
entire probability distribution of an
ensemble of random matrices, the formula (\ref{P2})
looses its meaning for non integer values of the parameter
$\beta$, characteristic to transition between the canonical ensembles.
However, in order to get possible interpolating formulae
for distribution of transmission coefficients
we will eventually allow $\beta$ to take any real value in $[0,2]$.

Relation (\ref{P2}) allows one to write the transmission coefficient
(\ref{T2}) in a simplified form
\begin{equation}
 T={M \over 2} -{1\over 2} {\rm Re[tr} (U_{\beta})],
\label{T3}
\end{equation}
and to obtain the distributions $P(T)$
by averaging above formula over
an appropriate ensemble of $M\times M$ unitary matrices $U_{\beta}$.

\section{Expectation values}
Since $\langle {\rm tr}(U_{\beta})\rangle =0$ for any $\beta$ one
obtains
\begin{equation}
 \langle T\rangle_{\beta}={M \over 2}.
\label{T4}
\end{equation}

Let ${\rm tr}(U_{\beta})=z=a e^{i \vartheta}$.
In a recent work \cite{HKSSZ96} the following average was derived
for canonical circular ensembles:
 $\langle a^2\rangle_{\beta}=2M/(2+\beta(M-1))$. Because the distribution of
phases
$\vartheta$ was shown to be uniform, the variances of both parts are
equal:  var(Re($z$))=var(Im($z$))=var($a)/2$. Using
(\ref{T3}) we get directly
\begin{equation}
 {\rm var}(T)_{\beta}=\langle (T-\langle T\rangle _{\beta})^2\rangle =
         {M \over 8+ 4 \beta(M-1)}.
\label{var1}
\end{equation}
Observe that for $\beta=2$ the variance equals to $1/8$, independently
of the matrix size M, while for
$\beta=1$ one has var($T)_1=M/4(M+1)$, in accordance with earlier
results \cite{GMMB96}. The variance growths with a decreasing $\beta$
and tends to $M/8$ in the limiting case $\beta\to 0$.

\section{Distribution of transmission coefficient $P_{\beta}(T)$ }

\subsection{The case M=1 }
For M=1 the "one dimensional matrix" $U=e^{i\varphi_1}$ and the phase
$\varphi_1$ is uniformly distributed in $[0,2\pi)$ for any
 ensemble. The variable $t=\cos(\varphi_1)$ has thus
the distribution
$P_t(t)=1/[\pi \sqrt{ (1-t^2)}]$. The transmission coefficient equal in
this case $T=1/2-t/2$ is therefore distributed according to
\begin{equation}
 P_{\beta,1}(T) = {1 \over \pi \sqrt{T(1-T)} }
\label{PT1}
\end{equation}
for any value of $\beta$.

\subsection{The case M=2 }
Let us start deriving the distribution $P_{\beta}(a)$ of the absolute
value of trace $a=|$Tr$(U)|$. For $M=2$ the JPD (\ref{P1})  reads
\begin{equation}
P_{\beta}[\varphi_1, \varphi_2 ]=C_{\beta,2} (\sin
 \phi )^{\beta},
\label{P3}
\end{equation}
where $\phi=(\varphi_1-\varphi_2)/2$.
The module of the trace $a=|1+e^{i2\phi}|=2\cos \phi$, so employing
(\ref{P3}) we obtain the required distribution
\begin{equation}
P_{\beta}(a) = c_{\beta} (4- a^2)^{{\beta-1\over 2}},
\label{Pa2}
\end{equation}
where $a\in [0,2]$ and the normalization constant reads
$$ c_{\beta} = {2^{1-\beta} \Gamma({\beta+2\over 2}) \over
   \sqrt{\pi} \Gamma({\beta+1 \over 2}) }. $$
In particular this distribution is flat for COE, while for CUE one gets
a semicircle and recovers the result already mentioned in
\cite{HKSSZ96}.  According to (\ref{T3}) the transmission coefficient
equals $T=1-(a \cos\vartheta) /2$. Due to rotational symmetry of the
ensembles the distribution of phase $\vartheta$ is uniform in $[0,2
\pi)$, so the distribution of $t=\cos\vartheta$ is
$P_t(t)=1/[\pi\sqrt{(1-t^2)}]$. Denoting $x=at$ and using (\ref{Pa2})
one writes an integral for the distribution of $x$
\begin{equation}
P_{\beta}(x) = {c_{\beta} \over \pi}
\int_x^2{(4-a^2)^{{\beta-1\over 2}} \over \sqrt{a^2-x^2} }da ,
\label{Pa3}
\end{equation}
which can be computed numerically in the general case.
By a linear change of variables $T=1-x/2$ it
provides the required distribution of transmission
coefficient $P_{\beta}(T)$. Moreover, for most interesting integer
values of $\beta$
 the above integral can be evaluated  analytically  giving

\begin{equation}
\label{PT2}
%P_{0}(T) = {2  \over \pi^2} K[\sqrt{ T(2-T)}]
%P_{1}(T) = {1  \over \pi} \ln {1+\sqrt{ T(2-T)} \over |1-T|}
%P_{2}(T) = {1  \over \pi} T(2-T) F({1\over 2},{3\over 2};2;T(2-T))
P_{\beta,2}(T) =  \left\{ \begin{array}{lll}
 {2  \over \pi^2} K[\sqrt{ T(2-T)}]
  &~\hbox{for}~& \beta=0, ~~(CPE);  \nonumber \\
{1  \over \pi} \ln {1+\sqrt{ T(2-T)} \over |1-T|}
  &~\hbox{for}~& \beta=1, ~~(COE);  \\
{4  \over \pi^2} T(2-T) D[\sqrt{T(2-T)}]
  &~\hbox{for}~& \beta=2, ~~(CUE),  \nonumber \\
\end{array}
\right.
\end{equation}
where $K$ and $D$ stand  for complete elliptic function of the first and
the third kind, respectively \cite{grad}. Somewhat more complicated
derivation of this formula in the case of COE and
CUE has already been given in \cite{GMMB96}, where the elliptic
function was expressed by the hypergeometric function as
  $D(k)={\pi \over 4} F({1\over 2},{3\over 2};2;k^2)$.

\begin{figure}
\unitlength 1cm
\begin{picture}(8,10)
\put(-1,-1){\includegraphics{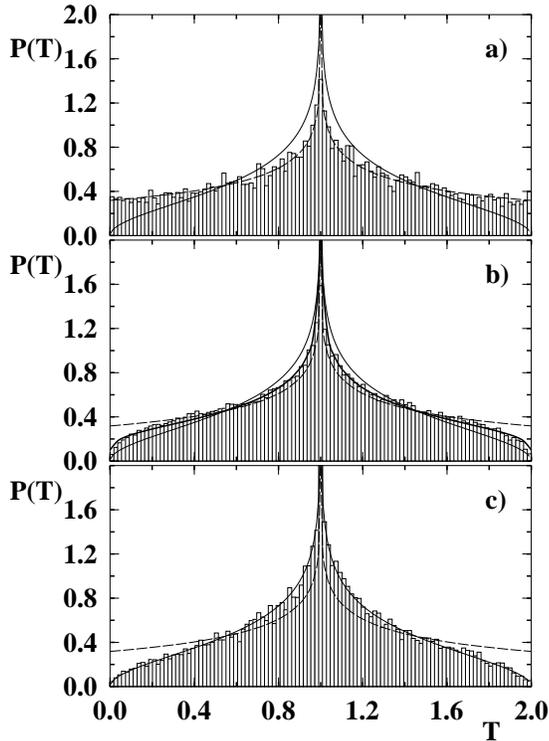}}
 \end{picture}
\unitlength 1bp
\caption{Distribution of the transmission coefficient $P(T)$ for
Poisson - orthogonal transition for $M=2$ obtained of $10^5$ random
matrices. The control parameter
$\delta$ is equal to a) $0.0$ (CPE); b) $0.4$; and c) $1.0$ (COE).
Solid and dashed narrow lines represent COE and CPE results,
respectively, while the bold
 line in Fig. b) stands for the best fit with $\beta=0.38$.}
\label{kafig1} \end{figure}

We have generated numerically random unitary matrices of interpolating
ensembles using the method proposed in \cite{ZK96}. Figure \ref{kafig1}
shows
the distribution of transmission coefficients $P(T)$ for $M=2$ and the
Poisson-COE transition for three different values of the control
parameter $\delta$.  Two narrow lines represent the
formula  (\ref{PT2}) with $\beta=0$ and $\beta=1$, characteristic for
the limiting cases of CPE and COE.
The bold line in the figure b), obtained in the interpolating case,
represents the best fit of (\ref{Pa3}) with $\beta=0.38$
(a coincidence between the values of
$\delta$ and $\beta$ is accidental). Good quality of the fit
reveals a certain validity of this
formula with non-integer values of $\beta$ for
ensembles in between the usual universality classes.

\subsection{Distribution $P_{\beta}(T)$ for large M}

Since the transmission coefficient $T$ is a function of $M$ random
variables with finite variances,
 one expects its distribution to be
Gaussian for large $M$.
 This fact can be proved rigorously for CUE. For this case
the traces $z= {\rm{tr}}(U)$ are distributed (in the
limit $M\to \infty$) as isotropic complex Gaussian variable
\cite{HKSSZ96}, what  guarantees
the Gaussian distribution as well for $x=$Re$(z)$, as well for
$T=(M-x)/2$. The Gaussian property also holds in Poissonian case,
for which  $x$
is the sum of $M$ independent terms, each being a cosine of uniformly
distributed random phases.

\begin{figure}
\unitlength 1cm \begin{picture}(8,10)
\put(-1,-1){\includegraphics{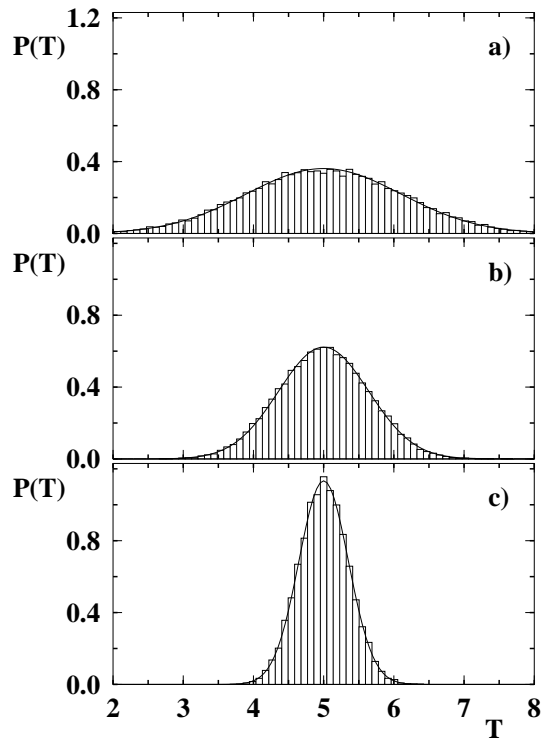}}
\end{picture}
\unitlength 1bp
\caption{As in Fig. 1
for Poisson - unitary transition and  $M=10$.
Solid lines represent the Gaussian distribution with $\langle
T\rangle=M/2$ and the variance given  by
(\protect\ref{var1}) for $\beta=0$ and $\beta=2$ and found
numerically for $\beta=0.4$. }
\label{kafig2}
\end{figure}

For $M>>1$ we conjecture the Gaussian distribution of $P(T)$ for any
value of $\beta$ with the mean equal to $M/2$ and the variance given by
(\ref{var1}). Numerical tests revealed Gaussian character of
this distribution for any $\beta$ already for $M=10$.
Figure \ref{kafig2} presents the distribution
of transmission coefficient for the Poisson-CUE transition and three
values of the control parameter $\delta$.

\section{Concluding remarks}

Composed ensembles of random unitary matrices where applied
for analysis of transmission coefficient
in symmetric chaotic systems described by S-matrices with block
symmetries. We simplified the derivation of
distribution $P_{\beta}(T)$ presented by Gopar et al.
\cite{GMMB96} and generalized
their results proposing an family of
interpolating distributions.

 It should be noted that the Poissonian case
$\beta=0$ discussed above is not capable to describe an effect of
localization. It corresponds rather to the case of scattering on a half
transparent mirror, at which each scattering mode acquires  a random
phase shift. Another generalization of the model designed to
describe effects of localization
is a subject of a subsequent publication \cite{szz99}.

It is a pleasure to thank Marek Ku{\'s}, Marcin Po{\'z}niak, Petr {\v
S}eba and Jakub Zakrzewski
for fruitful discussions. Financial support by Komitet Bada{\'n}
Naukowych is gratefully acknowledged.

\end{document}